\documentclass[trackchanges,twocolumn]{aastex7}
\usepackage{hyperref}
\usepackage{amsmath}

\newcommand{\Msun}{M$_\odot$} 
\newcommand{\Mstar}{M$_\star$}
\newcommand{\Mbh}{$M_\textrm{BH}$} 
\newcommand{\Lbh}{$L_\textrm{BH}$} 
\newcommand{\mesa}{\texttt{MESA}} 
\newcommand{\mesastar}{\mesa~\texttt{star}} 

\def\eg{\emph{e.g.}} 
\def\ie{\emph{i.e.}} 

\newif\ifref
\reftrue
\reffalse
\definecolor{darkred}{rgb}{0.75, 0, 0}
\newcommand{\mb}[1]{\ifref\textcolor{darkred}{#1}\else #1\fi}

\begin{document}

\title{MESA-QUEST: Tracing the Formation of Direct Collapse Black Hole Seeds via Quasi-stars}

\accepted{2026, Jan. 8}

\author[orcid=0009-0007-4394-3366, gname='Andy', sname='Santarelli']{Andrew D. Santarelli}
\affiliation{Department of Astronomy, Yale University, New Haven CT 06511}
\affiliation{Department of Physics, Illinois State University, Normal, IL 61761}
\email[show]{andy.santarelli@yale.edu}  

\author[orcid=0009-0002-5773-3531,gname=Claire,sname=Campbell]{Claire B. Campbell}
\affiliation{Department of Physics, Illinois State University, Normal, IL 61761}
\email{cbcamp1@ilstu.edu}

\author[orcid=0000-0002-5794-4286,gname=Ebraheem,sname=Farag]{Ebraheem Farag}
\affiliation{Department of Astronomy, Yale University, New Haven CT 06511}
\email{ebraheem.farag@yale.edu}

\author[orcid=0000-0003-4456-4863,gname='Earl', sname='Bellinger']{Earl P. Bellinger} 
\affiliation{Department of Astronomy, Yale University, New Haven CT 06511}
\email{earl.bellinger@yale.edu}

\author[orcid=0000-0002-5554-8896,gname='Priyamvada', sname='Natarajan']{Priyamvada Natarajan} 
\affiliation{Department of Astronomy, Yale University, New Haven CT 06511}
\affiliation{Department of Physics, Yale University, New Haven CT 06511}
\affiliation{Yale Center for the Invisible Universe}
\affiliation{Black Hole Initiative, Harvard University, Cambridge, MA 02138}
\email{priyamvada.natarajan@yale.edu}

\author[orcid=0009-0006-2122-5606,gname='Matt', sname='Caplan']{Matthew E. Caplan} 
\affiliation{Department of Physics, Illinois State University, Normal, IL 61761}
\affiliation{Department of Physics, University of Illinois Urbana-Champaign, Urbana, IL 61801}
\email{mecapl1@ilstu.edu}


\begin{abstract}
The origin of the first supermassive black holes (SMBHs) observed at redshifts $z\geq 9$ remains one of the most challenging open questions in astrophysics. Their rapid emergence suggests that massive ``heavy seeds'' must have formed early, possibly through the direct collapse of pristine gas clouds in the first galaxies. We present MESA-QUEST, a new framework built upon the Modules for Experiments in Stellar Astrophysics (MESA) code, designed to model the structure and evolution of quasi-stars—massive, radiation-supported envelopes hosting accreting black holes at their cores—believed to be the progenitors of direct-collapse black hole (DCBH) seeds. Our implementation introduces flexible boundary conditions representing both Bondi accretion and saturated-convection regimes, and explores the impact of several stellar wind and mass-loss prescriptions, including Reimers, Dutch, and super-Eddington radiation-driven winds. We find that quasi-stars can grow central black holes to $\geq 10^3$~M$_{\odot}$ under favorable conditions, with saturated-convection models yielding BH-to-total mass ratios up to 0.55M$_*$—five times higher than Bondi-limited cases. However, strong radiation-driven winds can dramatically curtail growth, potentially quenching heavy-seed formation unless balanced by sustained envelope accretion. Our results delineate the physical limits under which quasi-stars can remain stable and produce heavy seeds capable of evolving into the earliest SMBHs detected by JWST and Chandra. Future extensions will incorporate rotation, magnetic fields, and GR-radiation hydrodynamics to refine accretion physics and constrain the viability of the quasi-star pathway for early SMBH formation.
\end{abstract}


\section{Introduction}\label{sec:intro}

Most local galaxies appear to harbor a central supermassive black hole (SMBH). Dynamically determined masses of these central black holes reveal correlations with several key properties of their host bulges: masses, velocity dispersion and luminosities \citep{FerrareseMerritt2000,Tremaine+2002,2013ARA&A..51..511K} suggesting that the SMBHs and the hosts co-evolve. Recent JWST observations show that many massive galaxies formed in the universe by ${z \gtrsim 10}$ (or $\lesssim 500$~Myr after the Big Bang), some populated with central black holes and many active galactic nuclei (AGN) are in place by $z \gtrsim 7$ \citep{Castellano_2022, Naidu_2022, Harikane_2023, Labb__2023}. Even further, observations suggest that actively accreting SMBHs with masses $\gtrsim 10^{7-8}$~\Msun\ exist in high redshift galaxies such as UHZ1 \citep{Natarajan2024}. This early formation, in tandem with the galaxies' high SMBH to stellar mass ratio of $\sim 10\%$, has been argued to provide evidence for the ``direct collapse'' scenario of SMBH formation \citep{Natarajan2024,Whalen2023,coughlin2024,Natarajan_2017}. This mechanism predicts that heavy intermediate-mass black hole (IMBH) seeds form from the monolithic gravitational collapse of massive low-metallicity gas clouds in the cores of proto-galaxies \citep{Haehnelt1993, loeb1994, eisenstein1995, Madau_2001, bromm&loeb2003, begelman2006, lodato2006, alexander2014, Natarajan_2017}. 

Given that BHs are expected to grow via mergers and accretion over cosmic time \citep{HaehneltPNRees1998,Pacucci_2020_growth}, information about their initial seeding is expected to be erased. Therefore, directly accessing the highest redshift population offers the best prospects \citep{RicartePN2018, Pacucci_Loeb_2022} to constrain BH seeding models. The origin of SMBHs in the Universe remains an open question at the frontier of astrophysics research. High-redshift data from JWST are providing new insights with the detection of a handful of $z>9$ quasars that lie close to the seeding epoch (see for e.g. \cite{Maiolino+2024,bogdan2023+,Goulding+2023,kovacs+2024}. 

The inferred masses of some of these earliest JWST SMBHs detected at $z>9$ are in the range of $10^{6-7}$~\Msun; while those estimated using broad-line spectroscopy from the Sloan Digital Survey (SDSS), for example, are of the order of $\sim 10^9$~\Msun\ at $z\sim 6-7$ \citep{fan+2023}. These estimates of SMBH mass have been used to place constraints on a combination of seeding and growth trajectories in these epochs \citep{Fan+2004,mortlock11,Natarajan2011,Volonteri_2012,Wu+2015,banados+2018}. 

A range of theoretical seeding prescriptions that operate as early as $z \sim 20-25$ classified broadly as “light” and “heavy” seeding models have been proposed as starting points to account for the formation of observed SMBHs \citep{Natarajan2011,Volonteri_2012,Woods_2018}. Light seeds are believed to be remnants of the first generation of stars, the so-called Population III stars, that result in the production of initial BH seeds with $10-100$~\Msun\ \citep{Madau_2001}. 
 
 Heavy seed models, on the other hand, propose the formation of seeds $10^{4}\,-\,10^{5}\,M_{\odot}$ in several possible ways. First, heavy seeds could result from the direct collapse of pre-galactic gas disks \citep{loeb1994,volonteri05,lodato2006,begelman06,LodatoPN2007}. There are other promising proposed pathways and sites to form DCBH seeds at early epochs (i) from halos with supersonic baryon streaming motions relative to dark matter \citep{Stacy+2011,Schauer+2017}; (ii) in highly turbulent halos that are now thought to create the first quasars \citep{Latif+2022} and (iii) from major mergers of massive galaxies \citep{Mayer+2023,Mayer_Bonoli2019}. An additional pathway involves rapid, amplified early growth of originally light seeds that may end up in conducive cosmic environments, such as gas-rich, dense nuclear star clusters \citep{AlexanderPN2014}. Rapid mergers of light remnants in the early nuclear star clusters could also lead to the formation of heavy seeds at high redshifts as proposed by \cite{Devecchi_2009} as well as the runaway collapse of nuclear star clusters as proposed by \cite{Davies+2011}. In addition to these more conventional theoretical seeding models, primordial black holes (PBHs, \citealt{PBHs_1971}) that form in the infant Universe have also been explored as potential candidates to account for the origin of initial seeds for SMBHs (see \citealt{Cappelluti+2022} and references therein), as well as dark stars that are postulated to be powered by dark matter self-annihilation in their cores \citep{Freese+2016}.

The final stages of the formation of the DCBHs have been explored in simulations up to the stage where supermassive stars (SMSs) are produced. Rapid baryon collapse at the center of atomically cooled halos has been demonstrated to produce SMSs \citep{Hosokawa+2013,Woods+2017,Haemmerle+2018,Herrington_2023} which are then predicted to collapse by post-Newtonian GR instability or by depletion of core fuel at the end of post main sequence burning to form DCBHs via the formation of quasi-stars \citep{Begelman_2008}. Simulations have yet to track the subsequent formation and growth of the BH seed an initio and the stellar component in the host galaxy.

In this paper, we explore in detail, for the first time, the detailed evolution of the
Quasi-Star stage using the MESA-QUEST code that was custom built for this purpose. The outline of our paper is as follows: in Sec.~\ref{sec:channels}, we provide an outline of BH seed formation models and their cosmological contexts, and focus on tracing the DCBH formation process via quasi-stars in Sec.~\ref{sec:DC}. Summarizing previous work on quasi-stars in Sec.~\ref{sec:prevQS}, we then present our new implementation in MESA --- which we call MESA-QUEST --- with details that permit tracking their evolution and the results therein in Sec.~\ref{sec:MESAQS}. We focus on the key issue, the implementation of winds and mass loss in Sec.~\ref{sec:winds}, and close with a discussion of our conclusions and prospects for future work.

\section{Channels for BH seed formation}\label{sec:channels}

The process by which supermassive black holes (SMBHs) form remains an open question. In principle, a stellar-mass black hole can reach $10^9$~\Msun\ within the first Gyr after the Big Bang under the assumption that it accretes at the Eddington rate consistently for the entire time \citep[e.g.,][]{Haiman_2001}. However, typical black hole formation through the collapse of a massive star releases large amounts of energy in the form of ultra-violet radiation, leaving the stellar remnant in a low-density environment and stifling its accretion rate for an extended period of time \citep{Alvarez+2009}. This suggests that an alternate channel for the formation of SMBH seeds is necessary.

The masses of the $z>9$ JWST SMBHs require heavy initial seeds, as argued for in the case of UHZ1 and GHZ9 \citep{Natarajan2024,kovacs+2024} and/or special growth conditions that allow extremely efficient and rapid accretion for GNz11 \citep{Maiolino+2024}. In particular, the simultaneous detection of UHZ1 by JWST and the Chandra X-ray telescope; its IR colors; and the ratio of IR flux to X-ray match predictions and provide compelling evidence for a heavy initial BH seed formed via the direct collapse of gas \citep{Natarajan_2017}. 

The key issue for the rapid growth of initial light BH seeds is the time crunch, which requires optimal growth conditions, such as abundant gas reservoirs to allow rapid accretion while suppressing feedback that could disrupt gas inflow \citep[e.g.,][]{Haiman_2001,Pacucci_2015_growth_efficiency}. Although these optimal conditions are more likely to be available at very early epochs, they are difficult to sustain continuously to grow seed BHs starting with masses of 10-100 \Msun\ to the final SMBH masses powering $z > 6$ quasars \citep{Park+2011,Pacucci+2017,Pacucci_2020_growth}. This motivated the exploration of alternate seeding prescriptions, particularly those that might produce more massive initial seeds ranging in mass from $10^{3-5}$~\Msun\ or those in which rapid amplification of accreted mass can be achieved by fine-tuning cosmic conditions \citep[see][for reviews]{Volonteri&Bellovary2012,Haiman2013,Natarajan2014, Woods_2018}. 

SMBH seeds are believed to have formed from pristine gas in the very early universe ($z\gtrsim 15$).  Since atomic hydrogen is the only available coolant, the resulting Jeans masses can be large. Initial numerical studies of the formation of the first stars indicated that the initial mass function was likely tilted high compared to star formation occurring locally \citep{BL_03,Abel+2002}. In more sophisticated recent simulations, fragmentation occurs readily, resulting in stars with lower mass than previously found \citep{Clark+2011,Greif+2012,Latif+2013,Hirano+2014,Stacy+2016}. Notably, the remnants of this first generation of stars are expected to result in early light black hole seeds with masses ranging from 10-100~\Msun. Although forming these light seeds from the remnants of the first stars seems natural in the early Universe, growing them to $\sim\,10^9\,{M_{\odot}}$ in less than a Gyr, to account for the SDSS quasars, requires exquisitely fine-tuning conditions for gas supply and feeding. Light seeds have been shown to accrete gas with lower duty cycles and at sub-Eddington rates due to the typical cosmic environments that might harbor them \citep{Inayoshi+2016,Pacucci+2017}.  Heavy seeds, meanwhile, produced from the direct collapse of pre-galactic gas disks in pristine halos were therefore proposed as an alternative path to account for the SMBHs powering the brightest and highest redshift quasars \citep{lodato2006,begelman06}. In this DCBH seed formation picture, pristine gas in early disks that can become dynamically unstable will rapidly accrete mass to the center as their angular momentum can be rapidly dissipated through the formation of non-axisymmetric structures like bars on the small scales. This scenario, expected to operate in preferentially low angular momentum halos and hence low spin disks, would then in turn grow massive central objects \citep{Oh_2002,bromm&loeb2003,lodato2006,begelman06}.  DCBH formation occurs when the fragmentation of these gas disks is prevented by the suppression of cooling. This is achieved by ensuring rapid dissociation of any molecular hydrogen that might form, via irradiation of Lyman-Werner photons from an external field \citep{Shang+2010,Agarwal+2013,Regan+2017,Visbal_2018}. \citet{lodato2006,LodatoPN2007} first demonstrated that connecting the larger scale cosmological context with the fate of these pristine gas disks could be used to derive global properties like the initial mass function for heavy DCBH seeds \citep{LodatoPN2007}. Radiative and hydrodynamical cosmological simulations are available to probe the availability of direct collapse seeding sites within the context of the standard paradigm for structure formation \citep{Agarwal+2013,Agarwal+2014}. These studies agree that seeding sites are common enough in the early universe to explain the existence of observed ($z\sim 6-7$) quasars, but maybe too rare to account for the existence of typical SMBH in a local $L_*$ galaxy \citep{Wise+2019,Habouzit+2021}. In addition to direct collapse of gas, heavy seeds can also form when an initially light seed under the right set of circumstances undergoes extremely rapid growth via super-Eddington accretion (in $\sim 10^6$ year or less) or when aided by mergers with other seeds \citep{volonteri05,Devecchi_2009,AlexanderPN2014}. Intermediate mass black hole seeds that result from the dynamical core collapse of the first star clusters could also be a channel for early seed formation \citep{Omukai+2008,Devecchi_2009,Stone+2017,Natarajan2021}. 

\section{Tracking the evolution of Direct Collapse}\label{sec:DC}

 There are a number of possibilities and pathways for direct collapse to manifest, with the primary requirement being that typical fragmentation of the pre-galactic gas cloud must be prevented (\ie, temperatures $\gtrsim 10^4$~K and metallicities $\lesssim 10^{-5}~Z_\odot$) \citep{begelman&rees1978, Oh_2002}. The most prominent coolant in these early proto-galactic gas clouds is molecular hydrogen ($\mathrm{H_2}$), which eventually causes this fragmentation. However, ionization by Lyman-Werner radiation \citep[\eg{}, from a neighboring star-forming galaxy or from stripped binary stars,][]{Ferrara2013, Dijkstra2014, agarwal2014, Gotberg2019} can cause dissociation of $\mathrm{H_2}$ molecules and thus prevent fragmentation. Other possible direct collapse scenarios include the removal of angular momentum from non-axisymmetric perturbations and gravitational torques that occur during collapse, as well as after mergers of rare, high redshift massive galaxies when gas inflows cause an unstable gas disk to form and funnel in enough gas to collapse \citep{Begelman_2009, Choi_2013, Choi2015, Fiacconi2015, Mayer+2023}. 

Another scenario results when nuclear fusion begins at the core of the collapsing gas cloud, forming a convective core surrounded by a convectively stable envelope in what is known as a supermassive star \citep{begelman2010}. Eventually, the core collapses due to GR instability and a stellar-mass black hole is formed in the core. Alternatively, fusion may never occur, as it can be quenched by the continuous rapid infall of material \citep{begelman2006, coughlin2024}. However, the heating and neutrino losses still result in the same fate: a stellar-mass black hole nested at the center of a massive, slowly rotating, accretion radiation-supported envelope of gas. Regardless of which of these processes occurs, this end result has become known as a \textit{Quasi-Star} \citep{begelman2006, Begelman_2008, begelman2010, ball2012quasistars, Fiacconi2015}. 

Motivated by the need to account for many physical mechanisms to decipher the viability of quasi-stars to form heavy seeds, we developed MESA-QUEST to model and track the evolution of quasi-stars using \mesa\ \citep{Paxton2011, Paxton2013, Paxton2015, Paxton2018, Paxton2019} as it is capable of modeling them in greater detail with modern equations of state and opacities, detailed treatment of convection and winds, strong numerical stability, and with easier implementation of alternate physics. In this work, we begin with a description of previous work on quasi-stars, followed by a brief introduction to \mesa. We then discuss our black hole implementation within \mesa\ and explore both of the aforementioned boundary conditions. Finally, we explore multiple wind schemes, including the Reimers and ``Dutch'' schemes, which are built directly into \mesa\, in addition to implementing the custom schemes from \cite{Dotan_2011} and \cite{Fiacconi2015}. 

These changes and any future work are publicly available on Zenodo with a living version available on GitHub.\footnote{Zenodo:~\dataset[10.5281/zenodo.18156852]{https://doi.org/10.5281/zenodo.18156852} \\ 
GitHub:~\href{https://github.com/andysantarelli/MESA-QUEST}{https://github.com/andysantarelli/MESA-QUEST}}

\section{Previous Work on quasi-stars}\label{sec:prevQS}

Quasi-stars were first proposed by \cite{Begelman_2008} where they modeled these objects analytically using loaded polytropes. It was suggested that quasi-stars could exist long enough for the central black hole to accrete enough material to form a heavy seed with masses $\sim 10^{4-5}$~\Msun\ by $z\gtrsim10$. The unique characteristic of these objects, and what makes them appealing as heavy-seed progenitors, is that the black hole is able to accrete at the Eddington limit of the entire object, not just the black hole itself. This extreme accretion rate is due to the formation of an accretion disk very near the black hole, as well as the convection that occurs just outside it. This then allows for much more efficient release of energy and an increase in the accretion rate of the black hole. \cite{begelman2006} suggested that after reaching a certain photospheric temperature, the quasi-star may eventually eject its envelope, imposing a limit on the attainable black hole mass based on the mass of the entire object. 

\cite{ball2012quasistars} numerically simulated quasi-stars using the Cambridge \texttt{STARS} code with a spherically symmetric Bondi-Hoyle accretion scheme and inner boundary condition. They found that there was a stronger limit on black hole growth dependent on the black hole to total mass ratio than suggested by \cite{begelman2006}, showing that this ratio could not exceed $\sim0.167$. They argue that this limit is similar to that of the Sch\"onberg-Chandrasekhar limit, though the exact physical explanation remains to be determined. 

\cite{coughlin2024} explored this limit further, suggesting that it is due to a non-physical formulation of the inner boundary condition. They modeled a deeper, maximally convective region well within the previous boundary, where energy is transported outward by strong convection. Their ``saturated-convection" boundary condition leads to a black hole to total mass ratio of $~0.6$.

Quasi-stars must remain stable long enough for the central black hole to accrete a sufficient amount of material to create a heavy seed. There are then a number of stability concerns when it comes to this sustainability, including the effects of surface winds. \cite{Dotan_2011} argued that radiation-driven winds are prominent enough to unravel the envelope on a shorter timescale than that of black hole accretion, stifling the black hole growth for all quasi-star masses below \textit{a few} $\times~10^5$~\Msun. \cite{ball2012quasistars} employed the Reimers wind scheme, a simple analytic wind based on the mass, radius, and luminosity of the star \citep{Reimers1975} and show that the winds have very little effect on the lifetime of the star and thus the final black hole mass. However, this scheme is likely not appropriate for the extreme conditions in supermassive or quasi-stars and was implemented as a first approximation. \cite{Fiacconi2015} suggested that taking into account a previously neglected advection term in the radiation-driven wind calculation used by \cite{Dotan_2011} significantly decreases the impact of the wind on the evaporation timescale of the quasi-star. 

\section{MESA Implementation of Quasi-Star evolution}\label{sec:MESAQS}

Modules for Experiments in Stellar Astrophysics (\mesa) \citep{Paxton2011, Paxton2013, Paxton2015, Paxton2018, Paxton2019, Jermyn2023} is a collection of open-source suites designed to tackle a wide range of problems in computational stellar astrophysics. In this work, we are particularly interested in the 1-D stellar evolution module, \mesastar. This module utilizes adaptive mesh refinement and advanced timestep controls to solve fully coupled structure and composition equations. \mesa\ also uses a series of independently constructed modules to provide equation of state, opacity, nuclear reaction rates, element diffusion date, and atmosphere boundary conditions.

\subsection{Quasi-stars in MESA}

In previous works, we developed modifications to \mesa\ in order to model the quasi-star scenario \citep{Campbell_2025}. The black hole was implemented through adjustment of the boundary conditions and by including a sub-grid model for the black hole accretion and luminosity. Due to the massive black hole at the center of a quasi-star, general relativistic effects become important. We account for this using the Tolman-Oppenheimer-Volkov correction -- a modification to the equation for hydrostatic equilibrium via a change to the gravitational constant $G$, which is recalculated within \mesa\ at every timestep. Additionally, we use the Ledoux criterion for convection along with a mixing length of $\alpha_{mlt}=2$ as has also been used in previous models for supermassive stars \citep{Paxton2011,Herrington_2023}. 

To create an initial model, we evolve a typical stellar model with an initial mass of $20$~\Msun\ and an envelope accretion rate of $0.1~\textrm{M}_\odot \textrm{yr}^{-1}$ until it reaches $10,000$~\Msun, which is the total object mass for our fiducial quasi-star models. This method is similar to those used in \cite{Herrington_2023} when modeling supermassive stars. From this initial model, we then implement the aforementioned methods to account for the central black hole. This approach to initialization is not always necessary but does not significantly alter the quasi-star structure or lifetime in an impactful way and can thus be used regardless of necessity. 

We will now discuss the details of the boundary condition modification and the addition of black hole accretion and luminosity.  

\subsection{Boundary Conditions}\label{sec:BCs}

We choose to adopt an inner boundary at the point where hydrostatic equilibrium is no longer maintained. Precisely where this point is located is not fully understood. A number of physical mechanisms are at play near the black hole, including the convection caused by a steep temperature gradient, as well as the rotation of the black hole and any subsequent jets that may also transport energy. The interplay of these and other physical mechanisms makes it difficult to determine exactly where the hydrostatic equilibrium breaks down. 

Typically, \mesa\ solves for interior boundary conditions given by

\begin{align}
    m(r=0) &= 0 \\
    l(r=0) &= 0 
\end{align}

\noindent where $r$ is the radial coordinate, $m$ is the mass contained within $r$, and $l$ is the luminosity through the sphere of radius $r$. Implementation of a black hole requires a modification to these boundaries as noted below:

\begin{align}
    m(r=R_0) &= M_\mathrm{BH} \\
    l(r=R_0) &= L_\mathrm{BH}
\end{align}

\noindent where $R_0$ is the chosen boundary condition for the black hole, \Mbh\ is the mass of the black hole, and \Lbh\ is the chosen prescription for the black hole accretion luminosity through the inner boundary at $R_0$. 

We implement a ``cavity'' mass to account for mass contained within $R_0$ beyond the black hole mass \Mbh. We use the density profile

\begin{equation}
    \rho(r) = \rho_0 \left( \frac{r}{R_0} \right)^{-\frac{n}{2}}
\end{equation}

\noindent where $\rho_0$ is the density of the innermost cell, and $n=1$ or $3$ according to the structure of the surrounding star and whether the angular momentum is transported inward or outward, respectively \citep{Quataert_2000}. We then solve the integral 

\begin{equation}
    M_{\mathrm{cav}} = \int_{R_S}^{R_0}4\pi r^2 \rho(r) \textrm{d}r
\end{equation}

\noindent assuming that the Schwarzschild radius $R_S$ is much smaller than $R_0$. Our mass boundary condition then becomes the following: 

\begin{equation}
    m(r=R_0) = M_\mathrm{BH} + M_\mathrm{cav}.
\end{equation}

Using these modifications to \mesa, we explore multiple inner boundary conditions and accretion/luminosity schemes for to explore the evolution of quasi-stars, the physics and motivations of which are discussed in the following sections.

\subsubsection{Bondi Radius}

The first inner boundary condition we adopt is the same as originally used by \cite{ball2012quasistars} -- the Bondi radius $R_B$. This is the radius at which the thermal energy of the particles equals that of the gravitational potential energy such that:

\begin{equation}
    \frac{1}{2} m c_s^2 = \frac{GmM_\mathrm{BH}}{R_B} \\
    R_B = \frac{2GM_\mathrm{BH}}{c_s^2}
\end{equation}

\noindent where $m$ is the particle mass, $c_s$ is the sound speed, $M_\mathrm{BH}$ is the black hole mass and $G$ is the gravitational constant. Anything within this radius will eventually fall into the black hole, although local fluid motions within may still move both toward and away from the black hole \citep{ball2012quasistars, Ball2012MNRAS}. 

\subsubsection{Saturated-Convection Radius}

Recently \cite{coughlin2024} proposed a new, tighter inner boundary condition for quasi-stars where convection transports energy outward at the local maximum efficiency. The result is a deeper hydrostatic region than obtained with the original Bondi conditions because the material drifts slowly inwards, compared to the free fall that the Bondi condition assumes. 

In order to implement this new boundary condition $R_i$, we solve the differential equation

\begin{equation}
    K_{\rm i}\frac{d}{d\xi}\left[\xi^{-2}\left(\frac{dm_{\rm i}}{d\xi}\right)^{1/3}\right] = -\frac{m_{\rm i}}{\xi^4}\frac{dm_{\rm i}}{d\xi}, 
\end{equation}

\noindent where

\begin{equation}
    K_{\rm i} = \left(\frac{L\sqrt{3}}{2\eta}\right)^{2/3}\left(\frac{r_{\rm i}}{M_{\rm i}}\right)^{5/3}\frac{1}{G}. 
\end{equation}

\noindent These equations use dimensionless parameters $\xi = r/r_{\rm i}$ and $m_{\rm i} = m(r)/M_{\rm i}$, where $M_{\rm i}$ is the mass contained within $r_{\rm i}$. In this case, $K_i$ is the only free parameter to solve saturated convection solutions, and therefore can be used to determine the inner radius given the luminosity and mass of the object \citep{coughlin2024}. Additionally, our density profile must now be $\rho \propto r^{-1/2}$ if convection is to transport all of the energy around the black hole, similar to a convection dominated accretion flow (CDAF) solution \cite{Quataert_2000}. We then adjust $M_\mathrm{cav}$ accordingly. 

\subsection{Accretion \& Luminosity}

We now discuss the accretion schemes we have implemented within \mesa\ for modeling quasi-stars. The defining characteristic of quasi-stars, and what makes them a viable heavy seed formation channel, is the fact that they are able to accrete at the Eddington rate of the entire object \citep{begelman2006}. This allows the seed to grow within the appropriate time constraints as set by the observation of the earliest SMBHs \citep{Castellano_2022, Naidu_2022, Harikane_2023, Labb__2023, Natarajan2024}. 

We first begin with the convection-limited Bondi accretion rate, discussed previously in \cite{ball2012quasistars}. This is given by:

\begin{equation}
    \dot M_\mathrm{BH} = 16\pi \eta \frac{1-\epsilon}{\epsilon}\frac{\rho}{c_s c^2 \Gamma} (GM_{\mathrm{BH}})^2
\end{equation}

\noindent where $c_s$ is the sound speed; $\Gamma$ is the adiabatic index; and $\eta$ and $\epsilon$ are the convective and radiative efficiencies, respectively, for which we take a fiducial value of 0.1 for both. We then find the luminosity boundary condition via:

\begin{equation}\label{eq:qs_L}
    L=\frac{\epsilon}{1-\epsilon}\dot M_{\mathrm{BH}}c^2.
\end{equation}

\noindent Using this accretion rate and luminosity along with the Bondi radius, we obtain qualitatively similar results to those found in Part 3 of \cite{ball2012quasistars} \citep{Campbell_2025}.

The exact accretion rate of a black hole inside of a quasi-star envelope is not fully understood, partially due to the fact that there are many physical mechanisms at play (\eg, angular momentum, magnetic fields, jets, photon trapping, etc.) that impact the evolution. With this in mind, and with the previous calculations showing that the fiducial accretion rate is of order the Eddington rate of the entire object, we also implement a rate that scales with said luminosity by some chosen constant $\alpha$. We therefore define a new luminosity boundary condition

\begin{equation}\label{eq:alphaEdd}
    L_0 = \alpha L_\mathrm{E} = \alpha 4\pi \frac{c}{\kappa} GM_\star 
\end{equation}

\noindent where $M_\star$ is the total mass and $\kappa$ is the opacity at the core of the quasi-star, and obtain the corresponding accretion rate using Eq.~\ref{eq:qs_L} such that 

\begin{equation}
    \dot M_{BH} = \frac{1-\epsilon}{\epsilon} \frac{4\pi}{\kappa c}\alpha G M_\star.
\end{equation}

\noindent Using this rate allows us to study the stability and final seed mass for a range of rates. Understanding the most realistic accretion rates will require GRRMHD simulations with various additional physics, that is beyond the scope of this work. 

\begin{figure*}
    \centering
    \includegraphics[clip, width=1.0\linewidth]{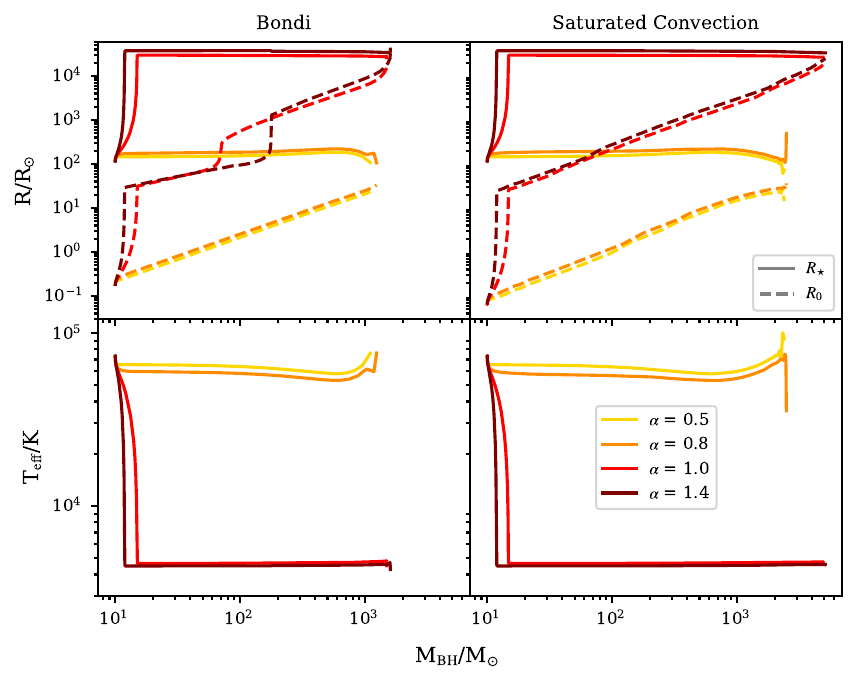}
    \caption{\textbf{The No Wind Scheme:} Total radius (top, solid) and inner boundary radius (top, dashed) as well as the effective temperature and core density (bottom) as a function of black hole mass for both the Bondi (left) and saturated-convection (right) inner boundary conditions without any wind scheme implementation. We include four models with various values of $\alpha$: $\alpha=0.5$ (yellow), $\alpha=0.8$ (orange), $\alpha=1.0$ (red), and $\alpha=1.4$ (maroon).}
    \label{fig:nowind}
\end{figure*}

\begin{figure*}
    \centering
    \includegraphics[clip, width=1.0\linewidth]{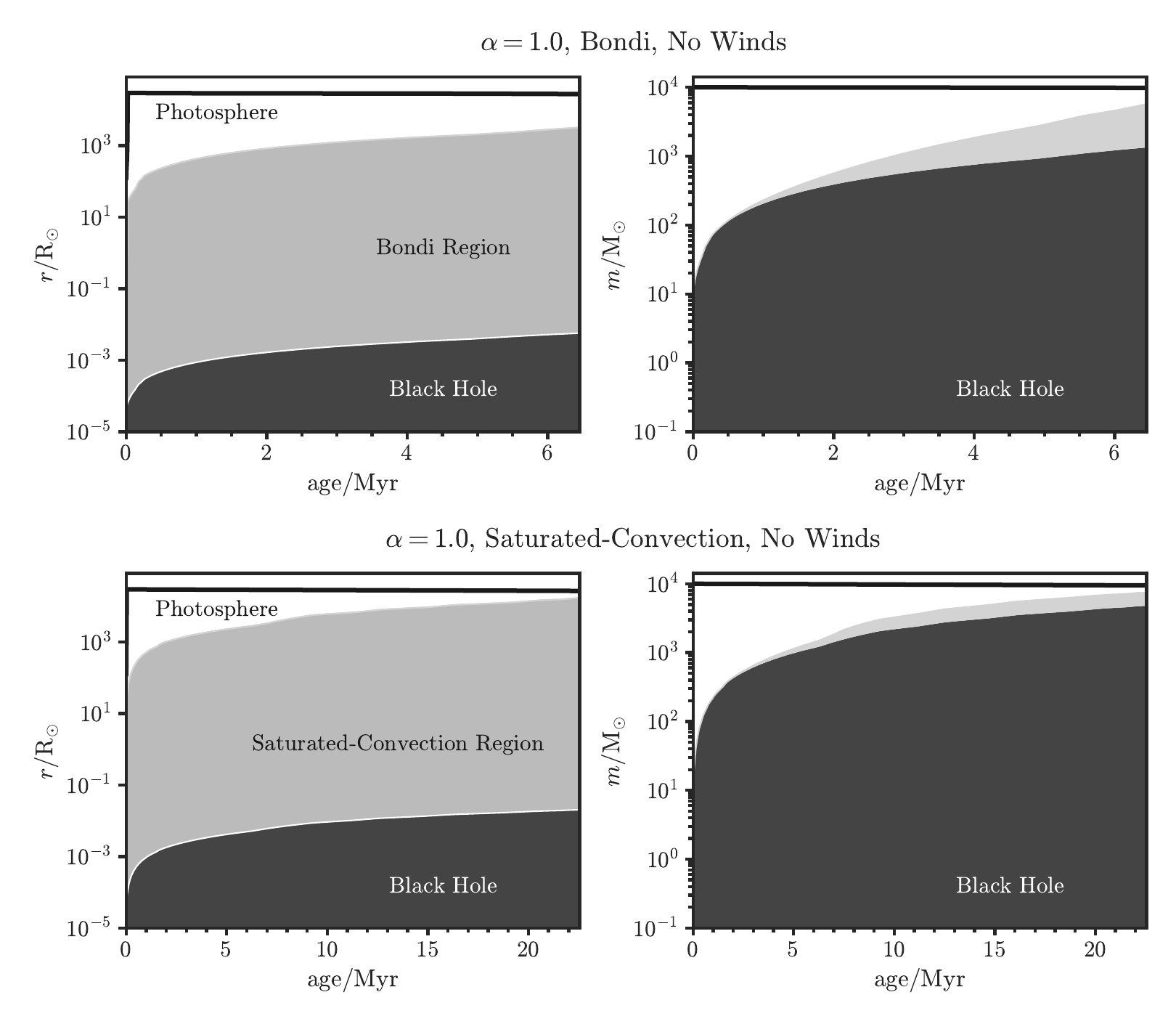}
    \caption{Kippenhahn diagrams showing the profile of the quasi-star without winds in both radial (left) and mass (right) coordinates for both the Bondi (top) and saturated-convection (bottom) models. We show the Schwarzschild region (black), boundary region (grey), envelope (white), and the photosphere (black line). Both models shown use $\alpha=1$.}
    \label{fig:kip_nowind}
\end{figure*}

Our fiducial models, shown in Figs.~\ref{fig:nowind} and \ref{fig:kip_nowind}, begin with a central black hole mass of 10~\Msun\ and a total mass of $10^4$~\Msun. We used a primordial composition of 75\% hydrogen and 25\% helium and chose $\alpha$ values of 0.5, 0.8, 1.0, and 1.4 to explore both potential regimes where the accretion rate is suppressed (\eg, by angular momentum) and where it is enhanced (\eg, by magnetic fields). 

\subsection{Results}

We observe similar final black hole masses to those seen in \cite{Ball2012MNRAS} and \cite{coughlin2024} for the Bondi radius and saturated-convection radius respectively, with a factor of $\sim5$ increase in the saturated-convection radius over the Bondi radius for $\alpha=1$ and $1.4$. Our models then also show that these mass limits are indeed not physical, but rather a limitation of the integration scheme -- the entire calculation, and thus the black hole growth, halts when the inner boundary reaches the surface of the quasi-star. 

Models with $\alpha < 1$ live longer, high-density lives and do not undergo the expected expansion that those with $\alpha \ge 1$ do. In fact, these quasi-stars remain relatively unchanged throughout their lifetimes. The accretion luminosity is not strong enough to extend the envelope, and the core opacity is high enough to prevent a runaway in the accretion luminosity otherwise seen in the beginning of higher $\alpha$ models. The quasi-star is also not hot enough to change ionization state, therefore keeping the opacity mostly constant and preventing a runaway in the luminosity. 

\begin{figure}
    \centering
    \includegraphics[width=1.0\linewidth]{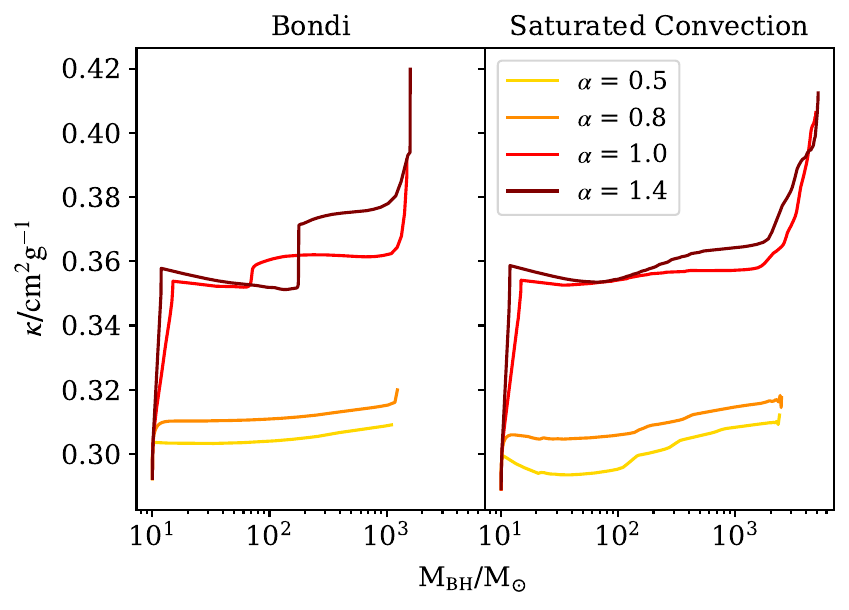}
    \caption{Core opacities for both Bondi (left) and saturated-convection (right) models. We include models with $\alpha=0.5$ (yellow), $\alpha=0.8$ (orange), $\alpha=1.0$ (red), and $\alpha=1.4$ (maroon).}
    \label{fig:opacities}
\end{figure}

Opacities for our fiducial models can be found in Fig.~\ref{fig:opacities}. We see that the scenario where $\alpha \simeq 1$ acts as a critical boundary between the aforementioned behaviors. The core opacity largely determines the bulk behavior of the accretion in the quasi-star. With the most substantial effects at higher BH masses near the end of the quasi-star's evolution, the opacity governs the evolution of the envelopes discussed in the following sections.

The saturated-convection models behave largely the same as the Bondi models, though the differences in final masses between the $\alpha$ values are much more evident. This appears to simply be due to the fact that the lower inner radius value allows the models that would grow larger to do so, while the models with $\alpha < 1$ have similar fates with little to no impact from the inner boundary. 

Of note is the step-like increase in the Bondi radius. This is likely due to the connection of convective regions that were previously separated by interstitial radiative regions \citep{ball2012quasistars}. This merging would cause the star to become fully convective, sharply decreasing the sound speed, thus increasing the Bondi radius. 

\section{Winds \& Mass Loss}\label{sec:winds}

Up to this point, our models have only included mass loss due to the conversion of a portion of the accreted mass to photons. However, mass loss due to surface winds is far more prominent and may occur at a much higher rate \citep{Dotan_2011}. Winds in ``typical'' stars are fairly well understood and have a wide variety of models for various star types \citep{Reimers1975, deJager1988, Nieuwenhuijzen1990, Bloecker1995, Nugis2000, Vink2001, vanLoon2005, Bj_rklund_2021}. In this section, we explore a few of these models that are built into \mesa\ and that best satisfy the conditions within quasi-stars as first-principle approximations. Additionally, we implement a super-Eddington radiation-driven wind scheme from \cite{Dotan_2011} that is specific to supermassive and quasi-stars. 

    \subsection{Reimers Scheme}

    Our first wind scheme is that of \cite{Reimers1975}, which was also implemented in \cite{ball2012quasistars} and will be referred to hereafter as the Reimers scheme. This scheme is commonly used for red giants and includes a scaling factor $\eta$ to allow for different mass loss efficiencies. The mass loss can be analytically calculated and is given by:

    \begin{equation}
        \dot M_\mathrm{wind} = 4\times 10^{-13} \times \eta \frac{LR}{M} 
    \end{equation}

    \noindent where $M$, $L$, and $R$ are the total mass, luminosity, and radius in solar units, respectively, and we use a typical value $\eta=0.5$ \citep{Reimers1975, Paxton2011}. 

    \subsection{``Dutch'' Scheme}

    Next we choose to explore a wind scheme used frequently for massive stars, defined as the ``Dutch'' scheme within \mesa. This scheme is a combination of the wind models from \cite{deJager1988}, \cite{vanLoon2005}, and \cite{Nugis2000} where the transitions between them occur based on the surface temperature. Specifically, the combination used is based on \cite{Glebbeek2009} where they formulated the scheme to work for stars with masses  $a~few\times 100$~\Msun. For surface temperature $T_\mathrm{eff} > 10^4$~K, the scheme of \cite{vanLoon2005} is used while that of \cite{deJager1988} is used for $T_\mathrm{eff} \leq 10^4$~K. It should be noted that the third model in the Dutch scheme of \cite{Nugis2000} is not used for our purposes, since it only occurs when the hydrogen mass fraction is $<0.4$. The temperatures and luminosities involved in the massive stars that the Dutch scheme was formulated for make this a better fit for quasi-stars, though it is likely still not appropriate. However, this scheme is built into \mesa\ and serves as an improvement on the Reimers scheme. 

    \begin{figure*}
        \centering
        \includegraphics[clip, width=1.0\linewidth]{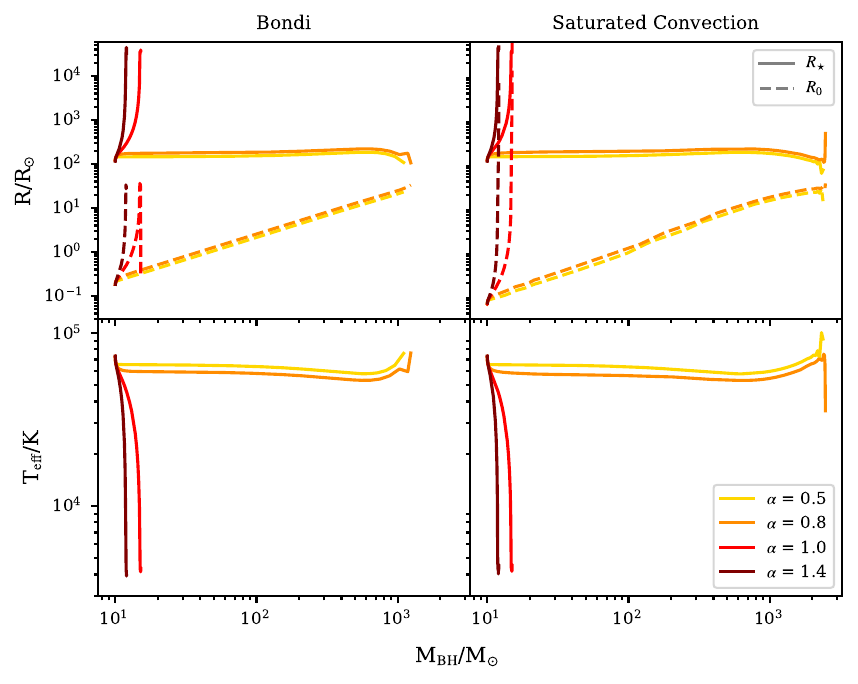}
        \caption{\textbf{Dutch Wind Scheme:} \textit{(See Fig.~\ref{fig:nowind})} Total radius (top, solid) and inner boundary radius (top, dashed) as well as the effective temperature and core density (bottom) as a function of black hole mass for both the Bondi (left) and saturated-convection (right) inner boundary conditions now with the Dutch wind scheme implemented. We include four models with $\alpha=0.5$ (yellow), $\alpha=0.8$ (orange), $\alpha=1.0$ (red), and $\alpha=1.4$ (maroon).}
        \label{fig:dutch}
    \end{figure*}

    \subsection{Super-Eddington Radiation-Driven Wind Schemes}

    Our final wind schemes come from \cite{Dotan_2011} and \cite{Fiacconi2015} which were derived specifically for quasi-stars. These schemes account for the effects of the super-Eddington luminosities and the accompanying outflows. \cite{Dotan_2011} calculates approximate hydrostatic solutions for the accretion powered envelope allowing for mass loss at the surface, resulting in a mass loss rate given by:

    \begin{equation}
        \dot M_\mathrm{wind} = 4\pi R_\star^2 \rho_\star \mu c_\star
    \end{equation}

    \noindent where $R_\star$, $\rho_\star$, and $c_\star$ are the radius, density, and sound speed at the surface, respectively. The addition of the Mach number $\mu$ allows the case where the sonic point (\ie, where the gas velocity is $c_s$) is not the critical point where the net forces balance each other (\ie, $R_\star$ when in equilibrium) \citep{Dotan_2011, Fiacconi2015}. 

    This scheme, however, was not modeled solving the full equations of motion and neglected the term for advection energy. \cite{Fiacconi2015} continued to numerically solve these equations with the inclusion of both diffusion and advection of energy. They found that the mass loss rate due to these winds can be reasonably well fit with power laws in the \Mbh-\Mstar\ plane using:

    \begin{equation}
        \dot M_\mathrm{wind} = 1.4 \times 10^{-4}~M_\star^{0.96}M_\mathrm{BH}^{0.17}~\mathrm{M_\odot yr^{-1}}.
    \end{equation}

    We implement this as our radiation-driven wind scheme within \mesa, with a few caveats. Firstly, this model does not take into account acceleration of the material when leaving the envelope. This acceleration occurs at a radius on the length-scale of the density scale height. However, on average this scale-height is of order $1\%~R_\star$, so can safely be ignored for this work. This scheme also does not account for the effects of rotation, which is outside the scope of this work \citep{Fiacconi2015}.

\subsection{Results}

     Implementing the Reimers wind scheme results in very little change to the final black hole mass when compared to our fiducial no-wind models. We see a $\sim$5-10\% decrease in mass for the $\alpha \geq 1$ Bondi models and $<$5\% in all saturated-convection models. This serves as a "best" case scenario for the quasi-star, placing a very small limit on the potential growth of the black hole. 

\begin{figure*}
        \centering
        \includegraphics[clip, width=1.0\linewidth]{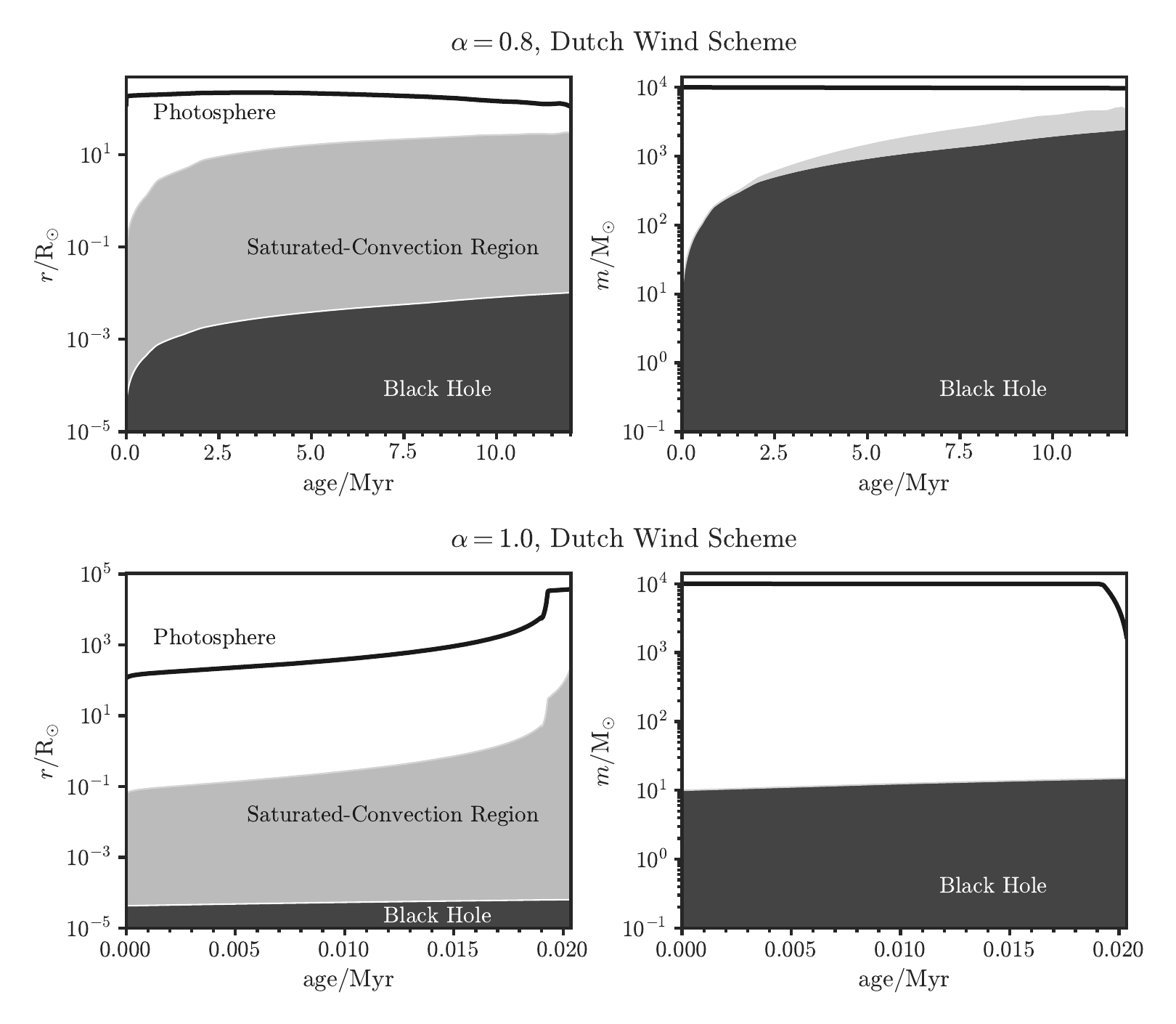}
        \caption{Kippenhahn diagrams showing the profile of the quasi-star using the saturated-convection boundary and including the Dutch wind scheme in both radial (left) and mass (right) coordinates. We show the Schwarzschild region (black), boundary region (grey), envelope (white), and the photosphere (black line) for models using $\alpha=0.8$ (top) and $\alpha=1.0$ (bottom).}
        \label{fig:kip_wind}
    \end{figure*}

     The results of implementing the Dutch wind scheme are plotted in Figs.~\ref{fig:dutch} and \ref{fig:kip_wind}. These winds are now strong enough to severely limit the final black hole masses. We see that the winds are enough to immediately eject the envelope upon expansion in the case of the $\alpha=1.4$ as it is enough to break the equilibrium that is otherwise reached. The $\alpha=1.0$ model lives only marginally longer, shedding its mass, while the black hole rapidly consumes what little remains. Our $\alpha < 1.0$ models remain largely unaffected due to both their lower accretion luminosity and their increased temperature. As stated previously, the transition between wind models for the Dutch scheme occurs at $T_\mathrm{eff} = 10^4$~K, which both low $\alpha$ models remain well above for the entirety of their lives. The Dutch scheme then serves as an intermediate case for a wind mass loss rate. 

    Implementing super-Eddington radiation-driven winds to our models results in significant mass loss for all values of $L_0$. These vigorous winds strip away the envelope before the black hole is able to accrete significantly, and the quasi-star only survives for $\sim 10^3$~yr. This scheme then serves as an upper end to our mass loss estimates. 

    \mb{In addition to these global mass loss schemes, it is also possible for a local region of the envelope's luminosity to exceed the Eddington limit and expel mass eruptively. This type of mass loss is supported by 3D radiation-hydrodynamic simulations and was also adapted to 1D for use within \mesa\ by \cite{Jiang_2015, Jiang_2018} and \cite{Cheng_2024} respectively. A study done in parallel to ours by \cite{hassan} explored this scheme further in the context of quasi-stars. They find a decrease in mass by a factor of $\sim 20$ in their equivalent of our $\alpha=1$ model. This sits within our range of wind schemes and represents a scenario between the Reimers and Dutch schemes. However, this process may be present alongside other wind drivers and become subdominant to the more severe mass loss rates of others.}
    
    It is of note that envelope accretion can heavily suppress the outflow and sustain the quasi-star's lifetime, allowing the black hole to accrete more of the envelope before it is shed \citep{Dotan_2011, Fiacconi2015, Haemmerl__2021}. This serves as motivation to pursue the addition of an envelope accretion scheme to our models in future work.

\section{Discussion \& Conclusions}

In this work, we have successfully implemented quasi-star models previously created by \cite{ball2012quasistars} using the Cambridge \texttt{STARS} code \citep{STARS} into \mesa\ \citep{Paxton2011}. This implementation includes the Bondi inner boundary condition as well as the Reimers wind scheme, both of which serve as our fiducial case. We also add the ability to take the accretion rate as a semi-free parameter in order to proxy the addition of other physics. In our fiducial models, we show that final black hole masses of $\sim 10^3$~\Msun\ can be attained even with accretion suppressing effects (\eg, angular momentum). 

We have implemented new inner boundary conditions based on a saturated-convection zone around the black hole proposed by \cite{coughlin2024}. Doing so allows the BH to grow to $\sim$0.55~M$_\star$, a factor of 5 increase from the previous $\sim$0.11~M$_\star$ achieved using the Bondi sphere as the inner boundary. The quasi-star, under these conditions, then becomes a heavy seed, capable of growing and reaching SMBH masses by early epochs thus explaining recent JWST observations. However, these results change dramatically when accounting for additional physics such as mass loss due to winds. 

We have studied three wind schemes in our models: the Reimers, Dutch, and super-Eddington radiation-driven winds. These serve as lower, middle, and upper bounds on mass loss rates for quasi-stars, as the actual values for surface mass loss are unknown without additional hydrodynamic and radiation transport simulations. We also presently do not account for envelope accretion, which is capable of balancing the mass loss and sustaining the envelope for an extended period of time. The Reimers and Dutch schemes may then serve as a proxy for that, given that the radiation-driven wind scheme is likely more realistic and catered to quasi-stars. We find that for the lower bound Reimers wind scheme, little change occurs to the final black hole mass for all accretion factors. \mb{In this case, the final SMBH masses will remain largely unaffected and are only limited by the size of the haloes that would would have formed the quasi-star in the first place.} The Dutch scheme, however, severely limits the growth of our higher $\alpha$ values, suggesting that if net mass loss occurs in this way that accretion enhancing effects (\eg, magnetic fields and jets) may actually limit the viability of quasi-stars to form heavy seeds. \mb{If $\alpha$ remains less than one, the final black hole mass is decreased by a factor of $\sim 2-3$, and would result in a lower upper limit on SMBH masses at a given redshift by a similar factor if we assume all heavy seeds are formed through this mechanism. When $\alpha \gtrsim 1$, the scenario is similar to that of a light-seeding mechanism; however, the rapid winds would leave the stellar mass remnant in an insufficient environment to accrete rapidly enough to become an SMBH by the appropriate redshifts. Quasi-stars can reasonably span a range of $\alpha$ values, which would result in both a decrease in the number density as well as the upper mass limit for SMBHs at a given redshift. The exact effect on the SMBH mass function across redshifts requires deeper population analysis and is outside the scope of this work.} Finally, the radiation-driven wind scheme presents the harshest limit on black hole growth, only allowing for a factor of two growth at most for all rates. If this wind is the most realistic, and accretion onto the envelope is not sufficient enough, this may severely limit quasi-stars as a potential heavy seed formation mechanism -- and thus an SMBH -- formation pathway. 

Our models are presently a first-principles approximation of quasi-star evolution and black hole seed growth. We are limited by our understanding of accretion physics as well as the modeling of the interplay between mass ejection and accretion at the surface of the quasi-star. We plan to explore this interplay further in future work in order to understand the range of black hole seed masses that can be attained and the relevant conditions. We also plan to explore new inner boundary conditions as there is no indication that either of the two presented here is more accurate than any other. We have also neglected the roles of rotation and magnetic fields on quasi-star evolution \mb{as they are non-trivial given both the 1D nature of \mesa\ and the need to implement coupled schemes for the envelope and BH.} We aim to explore the addition of these phenomena in future work.

Lastly, a better understanding of black hole accretion physics is vital to fully understanding quasi-star evolution and stability. Future work on GRRMHD simulations can provide more accurate rates and behaviors around the black hole and would allow us to more accurately study quasi-stars as a heavy seed formation channel.

\begin{acknowledgments}
This analysis contained within this paper was presented as a poster at the Simons Foundation SCEECS Annual Meeting in February 2025 and published as a Master's thesis in May 2025. Between then and the submission of this article we have learned of another group working on a similar problem whose work we would like to acknowledge here \citep{hassan}.  

Financial support for this publication comes from Cottrell Scholar Award \#CS-CSA-2023-139 sponsored by Research Corporation for Science Advancement. This work was supported by a grant from the Simons Foundation (MP-SCMPS-00001470) to M.C. This research was supported in part by the National Science Foundation under Grant No. NSF PHY-1748958. E.F. acknowledges support from the Yale Center for Astronomy and Astrophysics Prize Fellowship. P.N. acknowledges support from the Gordon and Betty Moore Foundation and the John Templeton Foundation that fund the Black Hole Initiative (BHI) at Harvard University where she serves as one of the PIs.
\end{acknowledgments}

\bibliography{mesa.bib, refs.bib}{}
\bibliographystyle{aasjournalv7}



\end{document}